\documentclass[aps,pra,twocolumn,floats,floatfix]{revtex4}
\usepackage{ifthen}
\usepackage{ifpdf}
\usepackage{color}

\ifpdf
\usepackage{graphicx}
\usepackage{epstopdf}
\else
\usepackage{graphicx}
\usepackage{epsfig}
\fi

\usepackage{latexsym}
\usepackage{amsmath}
\usepackage{amssymb}
\usepackage{bm}
\usepackage{wasysym}

\newcommand{\ei}{\hat{a}}
\newcommand{\eidag}{\hat{a}^{\dag}}

\newcommand{\Jx}{\hat{J}_x}
\newcommand{\Jy}{\hat{J}_y}
\newcommand{\Jz}{\hat{J}_z}

\newcommand{\beq}{\begin{equation}}
\newcommand{\eeq}{\end{equation}}
\newcommand{\beqa}{\begin{eqnarray}}
\newcommand{\eeqa}{\end{eqnarray}}
\newcommand{\la}{\langle}
\newcommand{\ra}{\rangle}

\newcommand{\hide}[1]{\textcolor{red}{[hidden]}}

\begin{document}

\title{Optimal Gaussian squeezed states for atom-interferometry in the presence of phase diffusion}

\author{Igor Tikhonenkov$^1$,  Michael G. Moore$^2$,  and Amichay Vardi$^{1,3}$}
\affiliation{$^1$Department of Chemistry, Ben-Gurion University of the Negev, P.O.B. 653, Beer-Sheva 84105, Israel\\
$^2$Department of Physics \& Astronomy, Michigan State Univerity, East Lansing, Michigan 48824, USA\\
$^3$ITAMP, Harvard-Smithsonian CFA, 60 Garden St., Cambridge, Massachusetts 02138, USA}

\begin{abstract}
We optimize the signal-to-noise ratio of a Mach-Zehnder atom interferometer with Gaussian squeezed input states, in the presence  interactions. For weak interactions, our results coincide with Phys. Rev. Lett. {\bf 100}, 250406 (2008), with optimal initial number-variance $\sigma_o\propto N^{1/3}$ and optimal signal-to-noise ratio $s_o\propto N^{2/3}$ for total atom number $N$. As the interaction strength  $u$ increases past unity, phase-diffusion becomes dominant, leading to a transition in the optimal squeezing from initial number-squeezing to initial {\it phase}-squeezing with $\sigma_o\propto\sqrt{uN}$ and  $s_o\propto\sqrt{N/u}$ shot-noise scaling. The initial phase-squeezing translates into hold-time number-squeezing, which is less sensitive to interactions than coherent states and improves $s_o$ by a factor of  $\sqrt{u}$.
\end{abstract}

\pacs{03.75.-b, 03.75.Lm, 03.75.Dg, 42.50.Xa}

\maketitle

\section{Introduction}
Heisenberg limited atom interferometers offer the possibility for compact, inexpensive measurement tools which will eventually operate at an unprecedented level of precision. Bose-Einstein condensates (BECs) in double-well potentials are well suited as a platform for such devices, as demonstrated by series of recent matter-wave interference experiments \cite{Andrews97,Shin04,Wang05,Albiez05,Jo07,JoChoi07,JoChoi07b,Schumm05,Hofferberth07,Est08,Boehi09,Gross10,Riedel10}. These experiments demonstrate that the double-well BEC system has the necessary phase-coherence, and a capacity for fine-tuning of the tunneling and interaction parameters, necessary to operate an atom interferometer at the maximum sensitivity allowed by quantum mechanics.

A  doube-well BEC interferometer is typically based on the Mach-Zhender Interferometer (MZI) paradigm, in which a bimodal input state is mixed by a 50/50 beam-splitter, then held for a fixed duration while the two modes acquire a relative phase differential $\theta$ (via an external field), and then mixed again by a second 50/50 beamsplitter. A measurement of the particle number-difference at the output then acts as an estimator for the accumulated phase differential. Aside from the preparation of the initial state, atom-atom interactions are typically neglected in theoretical treatments of the MZI.

If the input state is a two-mode coherent state (i.e. each particle is in the same single-particle orbital), the phase-estimation uncertainty, $\Delta\theta$ is governed by the Standard Quantum Limit (SQL), often referred to as the ``shot-noise-limit'', for which $\Delta\theta=1/\sqrt{N}$, with $N$ being the total number of particles used. In pioneering early work \cite{Cav81,Yur86,Hol93,Kitagawa93,Wineland94}, it became understood that quantum mechanics ultimately limits the phase-estimation uncertainty at the so-called Heisenberg limit $\Delta\theta= 1/N$, which is a factor $\sqrt{N}$ below shot-noise. In a typical realization, the best one can do is demonstrate ``Heisenberg-Limit scaling'',  which means that $\Delta\theta=q/N$, where $q$ is an $N$-independent constant. To reach the Heisenberg limit in a bimodal MZI, one must prepare a strongly number-sqeezed input state. While the maximally squeezed Twin-Fock state  (TFS), having exactly $N/2$ atoms in each mode, can achieve Heisenberg scaling at $\theta=0$, this requires two or three measurements \cite{Pezze08}, while for $\theta\neq0$, the TFS actually performs worse than shot-noise, and should thus be avoided.
As an alternative to the TFS, the Gaussian Squeezed State (GSS) with optimized squeezing exhibits a single-measurement phase-estimation uncertainty that smoothly approaches Heisenberg scaling as $\theta$ goes to zero \cite{Huang08}. 
Due to their strong interactions, Bose-condensed atomic vapors are ideal systems for creating  number-difference squeezed input states and using them as input into atom interferometers \cite{Est08,Boehi09,Gross10,Riedel10,Bou97,Eckert06,Sorensen01}. 
%The 50/50 beam-splitters is realized in double-well BECs, by temporarily lowering the tunneling barrier and allowing the two modes to mix for a fixed time interval.  
For these systems, the GSS is an excellent approximation to the ground-state at T=0, with the squeezing controlled by adiabatic variation of the interaction-to-tunneling ratio \cite{Imamoglu97}.

While strong interaction is essential for initial number-squeezed state preparation, it also limits the precision of the interferometer due to phase-diffusion during the phase acquisition time \cite{Castin97,Lewenstein96,Wright96,Javanainen97,Vardi01,Khodorkovsky08,Boukobza09,Tikhonenkov09, Grond10}. This process can be viewed  as the shearing of the initial phase-space distribution due to the different mean-field shifts experienced at different points in the distribution. Since the mean-field shift is proportional to the population imbalance between the two condensates, phase-diffusion is proportional to the relative-number variance, $\Delta$, during the hold time \cite{Jo07} with characteristic decoherence time of $1/(U\Delta)$, where $U$ is the interaction strength. Number-squeezed states ($\Delta\ll \sqrt{N}/2$) are transformed to phase-squeezed states($\Delta\gg\sqrt{N}/2$) by the first beam-splitter, thus providing sub-shot-noise accuracy, at the cost of increased sensitivity to phase-diffusion. By contrast, states which are number-squeezed during the phase-acquisition period, are far more robust, but suffer from inherently large readout uncertainty. The interplay between readout uncertainty and robustness against phase-diffusion implies that the initial squeezing should be optimized to give the best possible precision \cite{Grond10}.

It is conventional to describe the MZI as a device that measures $\theta$, the path-length difference between the two arms of the interferometer. While this is the proper way to view an optical interferometer, trapped-atom interferometers differ in that there is no fixed relation between time and distance. Thus we propose that the double-well condensate MZI be viewed as a device to measure the `bias', or energy differential, $\varepsilon$, between the two wells.  Unlike `flying particle' interferometers, the accumulated phase-shift $\theta=\varepsilon T$, in a `stationary particle' interferometer is not a fundamental measurable quantity, as the hold time $T$ is a free parameter, which can be used to optimize the measurement of $\varepsilon$. Here we use the freedom of the initial number-difference variance, $\sigma$, and the hold time, $T$,  to optimize the signal-to-noise ratio of a Mach-Zehnder atom interferometer with a GSS input. 

The optimization is performed first using exact numerical results, and then using approximate analytic expressions, with the two approaches showing good agreement. As the optimal performance improves with decreasing interaction strength, we assume that the experimenter has reduced the collision strength, $U$, as much as possible, given the constraints of the experimental set-up. Our goal is then to proscribe the optimal squeezing and hold-time based on this minimum value of $U$.  In the case where  $U\approx 0$ is obtained, e.g. via a Feshbach resonance, the experimental uncertainty, $\Delta U$ should be used in place of $U$.  

This approach connects previous work on optimizing a non-interacting interferometer \cite{Huang08} with the strong-interaction optimization of ``useful squeezing'' \cite{Grond10}, clearly showing a transition from optimal number-squeezing to optimal phase-squeezing, and mapping the transition region between the two regimes.

The two-site Bose-Hubbard model, initial state preparation, and optimization function are presented in Sect.~II. Numerical optimization results are described in Sect.~III and compared to analytic predictions in Sect.~IV. In Sect.~V we compare the best accuracy obtained for Gaussian states with that attainable from a coherent input, with conclusions presented in Sect.~VI.

\section{Model and initial preparation}
\begin{figure}[t]
\centering
\includegraphics[angle=0,width=0.5\textwidth]{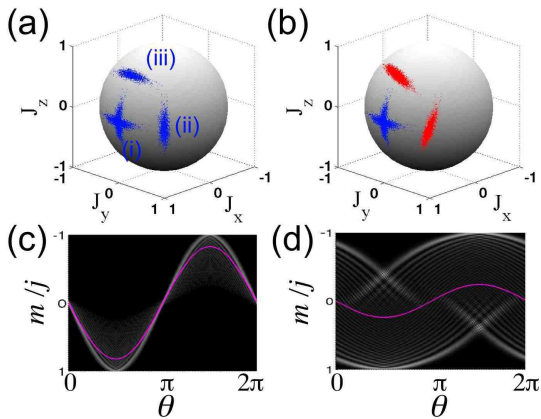}
\caption{(Color online) Mach-Zehnder interferometry with squeezed input in the presence of collision-induced phase diffusion. (a) The interferometer sequence on the Bloch sphere, without interactions. An initial number squeezed state is (i) rotated by $\pi/2$ about $J_x$ by the first beam-splitter, (ii) acquires a phase difference during the hold time $T$, and (iii) counter-rotated about $J_x$ by the second beam-splitter to give a final $Jz$ population imbalance readout. (b) Same with interactions. Phase diffusion results in the spreading of the squeezed states during the hold time, thus degrading the readout accuracy. (c) Final number distribution and average population imbalance (solid line) as a function of the acquired relative phase $\theta$ for a non-interacting gas. (d) same in the presence of phase diffusion, demonstrating reduced fringe visibility.}
\label{schematics}
\end{figure}

We consider a Mach-Zehnder interferometer, realized via the two-site Bose-Hubbard Hamiltonian \cite{Paraoanu01,Leggett01,Gati07}
\begin{equation}
\label{Ham}
H=-K\Jx+{\varepsilon}\Jz+U\Jz^2~.
\end{equation}
Here $K$, $\varepsilon$, and $U$ are coupling, bias, and interaction energies, where $U>0$ corresponds to repulsive interactions and vice versa.  The bias $\varepsilon$ may be positive or negative, depending on the energy detuning between the two modes. The SU(2) generators $\Jx=(\eidag_1 \ei_2+\eidag_2\ei_1)/2$,  $\Jy=(\eidag_1\ei_2-\eidag_2\ei_1)/(2i)$, and $\Jz=(n_1 - n_2)/2$, are defined in terms of the boson on-site annihilation and creation operators $\ei_i$, $\eidag_i$, with the conserved total particle number $n_1+n_2=N\equiv 2j$. The interferometer scheme (Fig.~\ref{schematics}(a)) consists of a fast $\pi/2$ beam-splitter rotation about $J_x$ (i), followed by relative-phase acquisition during a hold-time $T$ due to the bias detuning $\varepsilon$ (ii), and an opposite $\pi/2$ readout rotation about $J_x$ (iii). The final population imbalance $J_z^f$  is used to read the accumulated phase $\theta=\varepsilon T$ from which the bias, $\varepsilon$, is readily obtained.  

Assuming that the beam-splitter and read-out rotations are instantaneous with respect to the characteristic phase-diffusion time, the MZI can be described by the propagator
\beq
U_{MZI}(\theta, u,j)=e^{-i\frac{\pi}{2} \Jx }e^{-i\theta\Jz\left(1-(u/j)\Jz\right)} e^{-i\frac{\pi}{2}\Jx},
\label{UMZI}
\eeq
where $u=Uj/\varepsilon$.
This propagator acts on a Gaussian squeezed state, of the form,
\begin{equation}
|\sigma\rangle=\frac{1}{\sqrt{{\cal N}_\sigma}}\sum_{m=-j}^j \exp\left[-\frac{m^2}{4\sigma^2}\right]|j,m\rangle~,
\label{sigma}
\end{equation}
where $\sigma$ is the initial number-difference uncertainty, and ${\cal N}_\sigma=\sum_{m=-j}^j \exp(-m^2/(2\sigma^2))\approx\sqrt{2\pi}\sigma$.  
These states are an excellent approximation for the ground state of  Hamiltonian (\ref{Ham}) with $U>0$ and $\varepsilon\approx0$ \cite{Imamoglu97}, as well as to the dynamically squeezed states produced by single axis twisting of initially coherent states \cite{Kitagawa93,Gross10,Riedel10}, thus they can be readily generated with current experimental setups. 

Because the operators for phase-acquisition and phase-diffusion commute, the expectation value and variance  $\Jz$ at the output can be evaluated by applying the phase-diffusion operator to the  state vector, while incorporating the bias via a rotation of the observables by angle $\theta=\varepsilon \,T$. 
With the definitions
\beq
|\sigma,\theta\ra = e^{-i(u/j)\theta\Jz^2}e^{-i\frac{\pi}{2}\Jx}|\sigma\ra,
\eeq
and 
\beqa
\hat{J}_{z,\theta}&=&e^{i\theta\Jz}e^{i\frac{\pi}{2}\Jx}\Jz e^{-i\frac{\pi}{2}\Jx}e^{-i\theta\Jz}\nonumber\\
&=&\sin\theta\, \Jx -\cos\theta\, \Jy,
\eeqa
we see that $\la \Jz\ra_{out}=\la \sigma|U_{MZI}^\dag\Jz U_{MZI}|\sigma\ra$ can be expressed as
\beq
\label{Jz}
\la \Jz\ra_{out}=\la\sigma,\theta|\hat{J}_{z,\theta}|\sigma,\theta\ra=\sin\theta\, \la\Jx\ra_{\sigma,\theta},
\eeq
where
\beq
\la \hat{J}_\mu\ra_{\sigma,\theta}\equiv \la\sigma,\theta|\hat{J}_\mu|\sigma,\theta\ra;\quad\mu\in\{x,y,z\}.
\label{Jmu}
\eeq
The uncertainty in the final measurement is then given by
\beq
\label{DJzout}
\Delta J_{z,out}^2=\sin^2\theta\Delta J_{x,\sigma,\theta}^2+\cos^2\theta\Delta J_{y,\sigma,\theta}^2,
\eeq
where
\beq
\Delta J_{\mu,\sigma,\theta}\equiv \sqrt{\la \hat{J}_\mu^2\ra_{\sigma,\theta}-\la J_\mu\ra^2_{\sigma,\theta}}.
\label{DJmu}
\eeq 
We note that there are no terms proportional to $\la J_y\ra_{\sigma,\theta}$, $\la J_yJ_x\ra_{\sigma,\theta}$, and $\la J_xJ_y\ra_{\sigma,\theta}$ in Eqs. (\ref{Jz}) and (\ref{DJzout}) due to the symmetry of the state $|\sigma,\theta\ra$. 

It is useful to  define the bias-measurement signal-to-noise ratio (SNR) as
\beq
s\equiv\frac{\varepsilon}{\Delta\varepsilon}=\frac{\theta}{\Delta\theta}.
\eeq
From the error-propagation formula, 
\beq
\Delta\theta=\left[\frac{\partial\la J_z\ra_{out}}{\partial\theta}\right]^{-1}\Delta J_{z,out},
\eeq
it is then  straightforward to use (\ref{Jz}) and (\ref{DJzout}) to obtain
\beq
\label{SNRexact}
s=s(\sigma,\theta)=\frac{|\theta|\left|\langle\Jx\rangle_{\sigma,\theta}\right |}
{\sqrt{\Delta J_{y,\sigma,\theta}^2+\tan^2\theta\,\Delta J_{x,\sigma,\theta}^2}}.
\eeq
Because the state $|\sigma,\theta\ra$ depends only on the parameters $\{j,u,\sigma,\theta\}$, it follows that the optimal values, $(\sigma_0,\theta_0)$ that give the maximum SNR,  $s_0=s(\sigma_0,\theta_0$), as well as $s_0$ itself, are functions of $j$ and $u$ only.

%fig1
%%%%%%%%%%%%%%%%%%%%%%%
%%%%%%%%%%%%%%%%%%%%%%

\section{Numerical Results}

The goal of this paper is to find the optimal parameters, $(\sigma_o,\theta_o)$, that give the maximum signal-to-noise ratio, $s_o\equiv s(\sigma_o,\theta_o)$, for fixed $U$, $j$, and $\varepsilon$; and then use this  to determine the scaling of $s_o$ with particle number, $N=2j$, and interaction-to-bias ratio, $u$.  We note that in the case $s_o<1$, one should instead minimize the absolute uncertainty, $\Delta\varepsilon=\Delta\theta/T$. In this work, however, we will consider only the case $s_o\ge1$, with the ``minimum detectable bias'', $\varepsilon_{min}$ defined by $s_o(j,u=Uj/\varepsilon_{min})=1$.

\begin{figure}[t]
\centering
\includegraphics[angle=0,width=0.5\textwidth]{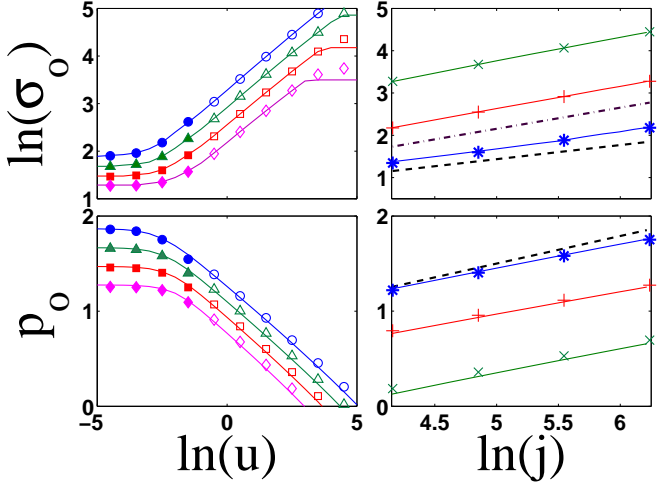}
\caption{(Color online) Optimal values of the initial relative-number variance $\sigma$ and precision $p$. Left panels depict  the dependence on the interaction parameter $u$ at  fixed atom number $j=64$ ($\diamond$), $128$ ($\square$), $256$ ($\triangle$), and 512 ($\circ$). Right panels show the dependence on $j$  at fixed $\ln(u)=-2.5$ (*), 0 (+), and 2.5 ($\times$).  Solid lines are exact values whereas symbols correspond to the weak-interaction estimate of Eqs.~(\ref{weak}) (filled) and the stron-interaction estimates of Eqs.~(\ref{strong}) (unfilled). Dashed lines denote the $j$ dependence for $u=0$, whereas the dash-dotted line marks the width of a spin coherent state, separating initial number-squeezing below it from phase-sqeezing above. The transition from number- to phase-squeezing takes place at $\ln(u)\approx 1$}
\label{GSSOPT}
\end{figure}

\begin{figure}[t]
\centering
\includegraphics[angle=0,width=0.5\textwidth]{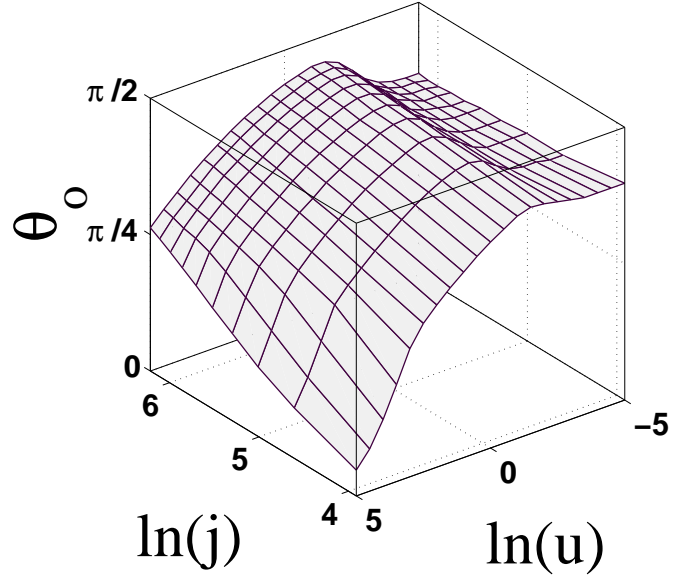}
\caption{(Color online) Optimal acquired phase $\theta_o$ for Gaussian squeezed states, as a function of the interaction parameter $u$ and the particle number $j$.}
\label{GSST}
\end{figure}

 The results of such optimization, using the numerical evaluation of Eq.~(\ref{SNRexact}) are shown in Fig.~\ref{GSSOPT} and Fig.~\ref{GSST}. In Fig.~\ref{GSSOPT} we plot the optimal squeezing $\sigma_o$, and the resulting maximized precision $p_o=\log_{10}s_o$ ($p_o$ directly corresponds to the number of significant figures of the read-out), as a function of the parameters $u$ and $j$. In Fig.~\ref{GSST} we plot the optimal acquired phase $\theta_o=\epsilon T_o$ as a function of $j$ and $u$. The fact that the acquired phase is optimized, rather than the hold time is discussed in detail in the next section. Symbols in Fig.~\ref{GSSOPT} correspond to the analysis presented in Section IV. The optimal initial number variance increases with increasing interactions, whereas precision is degraded. At the limit of small $u$, we obtain that $\sigma_o$ scales as $j^{1/3}$ and $s_o$ scales as $j^{2/3}$ (dashed lines in Fig.~\ref{GSSOPT}), in agreement  with Ref. \cite{Huang08}. As the interactions increase, these power laws are replaced by a $\sim\sqrt{j}$ dependence of both quantities. Most significantly, a transition from optimal initial number-squeezing ($\sigma_o<\sqrt{j/2}$) to optimal initial phase-squeezing ($\sigma_o>\sqrt{j/2}$) takes place as the interaction parameter $u$ crosses unity. This transition results from the interplay of projection noise minimization by initial number squeezing (i.e. hold-time phase-squeezing, resulting in a narrower 'phase-dial') and phase-diffusion control by initial phase-squeezing (i.e. hold-time number-squeezing, rendering the state more robust against phase-diffusion).  As the interactions increase, phase-diffusion determines the interferometer's precision and initial phase-squeezing is preferred.  With current experimental set-ups, number-squeezed states are more readily obtained than phase-squeezed states \cite{Est08}. As the first-beam splitter rotates a number-squeezed state into a phase-squeezed state, initial phase-squeezing in the MZI picture is obtained in practice by preparing a number-squeezed state with $\sigma\to\frac{j}{2\sigma}$, and eliminating the first `beam-splitter' $\Jx$-rotation, so as to have a number-squeezed state during phase-acquisition.

%fig2
%%%%%%%%%%%%%%%%%%%%%%%
%%%%%%%%%%%%%%%%%%%%%%

%fig3
%%%%%%%%%%%%%%%%%%%%%%%
%%%%%%%%%%%%%%%%%%%%%%

\section{Analytic optimization in the presence of Phase-diffusion}

%%%%%%%%%%%%%%%%%%%%%%%
Noting that $\la\Jy\ra_{\sigma,\theta}=0$ identically and using Eq.~(\ref{DJmu}), we may rewrite the signal-to-noise ratio of MZI output (\ref{SNRexact})  as,
\beq
s(\sigma,\theta)=\frac{|\theta|}{\sqrt{Q(\sigma,\theta)}},
\eeq
where 
\beq
Q(\sigma,\theta)=\frac{\la\Jy^2\ra_{\sigma,\theta}}{\la\Jx\ra^2_{\sigma,\theta}}+\tan^2\theta \left(\frac{\la\Jx^2\ra_{\sigma,\theta}}{\la\Jx\ra_{\sigma,\theta}^2}-1\right).
\eeq 
The optimized signal-to-noise ratio is then obtained by minimizing $Q(\sigma,\theta)$ with respect to $\sigma$ for fixed $\theta$ to obtain $\sigma_o(\theta)$, and then maximizing $s(\sigma_o(\theta),\theta)$ with respect to $\theta$.

In order to derive an approximate analytic expression for $Q(\sigma,\theta)$, we rely primarily on the approximation that a rotated Gaussian state is itself a Gaussian. Thus we make the anzatz,
\beq
|\sigma,\theta\ra=\frac{1}{\sqrt{2\pi}\Delta}\sum_{m=-j}^{j-1}|j,m\ra \exp\left[-m^2\left(\frac{1}{4\Delta^2}+i\frac{u\theta}{j}\right)\right],
\label{ansatz}
\eeq
for which $\Delta J_{z,\sigma,\theta}=\Delta$.  This variance is clearly a constant of motion during phase-acquisition (i.e. it is independent of $u$ and $\theta$). As the initial state $|\sigma\ra$ is a minimum uncertainty state with $\Delta J_z=\sigma$, it follows that the state immediately after the first $\Jx$-rotation, $|\sigma,\theta{=}0\ra$, is also a minimum uncertainty state, with $\Delta J_y=\sigma$. From the Heisenberg uncertainty principle, it then follows that
\beq
\Delta=\frac{\la \Jx\ra_{\sigma,\theta=0}}{2\sigma}.
\eeq
Noting that $\la \Jx\ra_{\sigma,\theta=0} =\la\sigma|\Jx|\sigma\ra=\la \sigma|\hat{J}_+|\sigma\ra$, with $\hat{J}_\pm=\Jx\pm i\Jy$, we arrive at
\beq
\Delta=\frac{e^{-\frac{1}{8\sigma^2}}}{2\sigma^2\sqrt{2\pi}}\sum_{m=-j}^j\sqrt{(j{-}m)(j{+}m{+}1)}e^{-\frac{(m+\frac{1}{2})^2}{2\sigma^2}}.
\label{Deltasum}
\eeq
Replacing the sum by an integral with respect to $x=(m{+}1/2)/j$ gives, 
\beqa
\Delta&\approx&\frac{j^2}{2\sigma^2}\frac{e^{-\frac{1}{8\sigma^2}}}{\sqrt{2\pi}}\int_{-1}^1 dx\, \sqrt{1-x^2} e^{-\frac{j^2}{2\sigma^2}x^2}\nonumber\\
&\approx&\frac{j}{2\sigma}e^{-\frac{1}{8\sigma^2}}\left(1-\frac{\sigma^2}{2j^2}-\frac{3\sigma^4}{8j^4}+\ldots\right),
\label{Delta}
\eeqa
where we have made the approximation $j+1/2\approx j$, and dropped terms proportional to $e^{-\frac{j^2}{2\sigma^2}}$.

To proceed further, make use of conservation law, $\Jx^2+\Jy^2+\Jz^2=J^2\approx j^2$, which together with the relation $\Jx^2-\Jy^2=\frac{1}{2}(\hat{J}_+^2+\hat{J}_-^2)$, gives us
\beq
2\la\Jx^2\ra_{\sigma,\theta}=j^2-\Delta^2+\mbox{Re}\left\{\la\hat{J}_+^2\ra_{\sigma,\theta}\right\},
\eeq
and
\beq
2\la\Jy^2\ra_{\sigma,\theta}=j^2-\Delta^2-\mbox{Re}\left\{\la\hat{J}_+^2\ra_{\sigma,\theta}\right\}.
\eeq
Thus in order to calculate $Q(\sigma,\theta)$, we need to compute only $\la \Jx\ra_{\sigma,\theta}$ and $\la\hat{J}_+^2\ra_{\sigma,\theta}$. Following the same procedure used to arrive at (\ref{Delta}), we find
\beqa
\la\Jx\ra_{\sigma,\theta}&\approx& j \exp\left[-\frac{1}{8\Delta^2}-\frac{2u^2\theta^2\Delta^2}{j^2}\right]\nonumber\\
&\times&\left(1-\frac{\Delta^2}{2j^2}-\frac{3\Delta^4}{8j^4}+\frac{2u^2\theta^2\Delta^4}{j^4}+\ldots\right),
\eeqa
and
\beqa
\la\hat{J}_+^2\ra_{\sigma,\theta}&=&j^2\exp\left[-\frac{1}{2\Delta^2}-\frac{8u^2\theta^2\Delta^2}{j^2}\right]\nonumber\\
&\times&\left(1-\frac{\Delta^2}{j^2}+\frac{16u^2\theta^2\Delta^4}{j^4}\right).
\eeqa
Combining these results, and expanding in terms of the three independent small parameters: $1/\sigma$, $\sigma/j$, and the phase-diffusion winding-angle, $\theta_d=u\theta/\sigma$; we arrive at $Q(\sigma,\theta)=Q_{NI}(\sigma,\theta)+Q_I(\sigma,\theta)$, where
\beq
Q_{NI}(\sigma,\theta)\approx\frac{\sigma^2}{j^2}+\tan^2\theta\frac{1}{32\sigma^4},
\label{QNI}
\eeq
gives the effects of squeezing in the absence of interactions, and
\beq
Q_I(\sigma,\theta)\approx\left(\frac{u\theta}{\sigma}\right)^2\left(1+\tan^2\theta\,\frac{\sigma^2}{j^2}\right)
\label{QI}
\eeq
determines the effects of interactions. 
The extremum condition, $\partial_{\sigma }Q(\sigma,\theta)=0$, can be expressed as
\beqa
x^3-bx-1&=&0,
\label{extremum}
\eeqa
where we have introduced the dimensionless parameters $x=(\sigma_o/\sigma_{NI})^2$,  $b=(u/u_c)^2$, and $\alpha=\sqrt[3]{\tan(\theta)/4}$, where
\beq
\label{sigmaNI}
\sigma_{NI}=\alpha\, j^{1/3},
\eeq
is the non-interacting solution \cite{Huang08}, and
\beq
u_c=\frac{\alpha^2}{\theta j^{1/3}},
\eeq
is the critical collision parameter above which interactions predominantly determine the optimal performance.

It is readily seen that in the non-interacting case ($b\to0$), we have $x=1$, in which case the optimized signal-to-noise ratio is
\beq
\label{sNI}
s_{NI}=\sqrt{\frac{2}{3}}\frac{\theta}{\alpha}\, j^{2/3}.
\eeq
The optimal angle $\theta_{NI}$ is then found via $(1-\theta\partial_\theta)\alpha(\theta)=0$, which has the solution $\theta_{NI}=1.14$ radians, giving $\sigma_{NI}=0.86\, j^{1/3}$, and $s_{NI}=1.14\, j^{2/3}$. While the non-interacting MZI with GSS input can indeed exhibit ``Heisenberg scaling'' at $\sigma=1$ and $\theta=1/j$, our optimization shows that it is more advantageous to instead use $\theta\sim 1$ and $\sigma\sim j^{1/3}$; a result that follows from the point-of-view that the property actually being measured is the bias, $\varepsilon$, rather than the phase, $\theta$. The Heisenberg-limited measurement gives $\Delta\theta=1/j$, but with a signal-to-noise ratio of $s\sim 1$, smaller than $s_o$ by a factor $j^{2/3}\gg 1$.   

Furthermore, this shows that in the non-interacting case, there is no minimum observable bias. A single measurement of any $\varepsilon$, no matter how small, can yield a maximum precision of $p_o\approx \frac{2}{3}\log_{10}j$ (meaning that increasing the atom number by $\times 30$ results in one additional decimal place of precision). In contrast, with no squeezing ($\sigma=\sqrt{j/2}$), the maximum obtainable precision scales as $\frac{1}{2}\log_{10}j$, and therefore requires an increase of $\times 100$ atoms for each additional significant figure. Of course in practice, at long hold-times, phase-diffusion inevitably degrades the performance, hence the optimized performance in the presence of interactions $(u\neq 0)$ is ultimately of more interest than the non-interacting case.

In the weakly interacting regime $b\ll 1$, Eq. (\ref{extremum}) can be written as 
\beq
x^3-1=bx.
\eeq
Treating the r.h.s. as a perturbation gives the solution
\beq 
x\approx 1+ \frac{b}{3},
\eeq
resulting in the weak-interaction behavior,
\begin{eqnarray}
\sigma_o&\approx&\sigma_{NI}\sqrt{1+\frac{1}{3}\left(\frac{u}{u_c}\right)^2}\\
\label{weak}
s_o&\approx&\sqrt{\frac{2}{3}}\frac{\theta_o}{\alpha}j^{2/3}\left[1+\frac{2}{3}\left(\frac{u}{u_c}\right)^2\right]^{-1/2}~.
\end{eqnarray}

The strong interaction regime $b\gg1$ was studied in Ref. \cite{Grond10}, using generic phase-space arguments for the dynamics rather than an explicit ansatz. The $\sigma^2/j^2$ and $u^2\theta^2/\sigma^2$ terms on the r.h.s. of  Eqs.~(23) and (24) are respectively identical to the noise and phase-diffusion terms  used in Eq.~(20) of \cite{Grond10} to minimize the  ``useful squeezing'', defined as  $\sqrt{2j}\Delta J_y/ |\langle\Jx(\sigma,\theta)\rangle |$ \cite{Kitagawa93,Wineland94}. However, assuming that the contribution of $\Delta J_x$  is negligible is only valid for initial preparations where neither $\Delta J_y^i$ nor $\Delta J_z^i$ significantly exceed the coherent-state variance $\sqrt{j}$. This greatly reduces the range of $\sigma$ available for optimization. In particular when $u=0$, ignoring the $\Delta J_x$ contributions to $Q(\sigma,\theta)$ fails to reproduce Eqs.~(\ref{sigmaNI}) and ({\ref{sNI}). In Sec. VI, we provide a detailed comparison of our results with those of Ref. \cite{Grond10} .

In the strongly-interacting regime, $b\gg 1$, Eq. (\ref{extremum}) can be expressed as
\beq
x^2-b=\frac{1}{x}.
\eeq
Again, treating the r.h.s. as a perturbation gives 
\beq
x=\sqrt{b}+\frac{1}{2b},
\eeq
resulting in
\beqa
\sigma_o&=&\sigma_{NI}\left(\frac{u}{u_c}\right)^{1/2}\sqrt{1+\frac{1}{2}\left(\frac{u_c}{u}\right)^3},\\
s_o&=&\frac{\theta_o}{\alpha}j^{2/3}\sqrt{\frac{u_c}{2u}}\left[1+\frac{1}{4}\left(\frac{u_c}{u}\right)^2\right]^{-1/2}.
\label{strong}
\eeqa
To leading order, this gives the strongly-interacting results, 
\beq
\sigma_o\approx \sqrt{u\theta_o j},\quad s_o\approx \sqrt{\frac{\theta_o j}{2u}}.
\eeq

The optimal acquired phase $\theta_o$ is obtained for weak and strong interactions respectively, by substitution of  $\sigma_o$ from Eq.~(\ref{weak}) and Eq.~(\ref{strong}) into $\partial(Q/\theta^2)/\partial\theta=0$. Consequently, for strong interactions ($u>u_c$),
\begin{equation}
2\theta_o^2\tan\theta_o\left( 1+\tan^2\theta_o\right)\approx j/u,
\label{thetastrong}
\end{equation}
whereas for weak-interactions ($u<u_c$) we have,
\begin{equation}
%\frac{2\theta}{\sin 2\theta}=C_0+2\frac{u^2\theta^2 j^{2/3}}{\alpha^4}~,
\sin 2\theta_o\left( 1-\frac{2}{3}\frac{u^2\theta_o^2 j^{2/3}}{\alpha^4 }\tan^2\theta_o\right)\approx\frac{2}{3}\theta_o~.
\label{thetaweak}
\end{equation}
%where $C_0=2\theta_{NI}/\sin2\theta_{NI}$, 
When $u=0$, Eq.~(\ref{thetaweak}) reduces to $\sin(2\theta_o)/(2\theta_o)\approx1/3$ , resulting in the appropriate non-interacting solution $\theta_o=\theta_{NI}$,. Eqs.~(\ref{thetastrong}) and (\ref{thetaweak}) are in good agreement with the numerical optimization shown in Fig.~\ref{GSST}. 

The weak-interaction Equations (\ref{weak}),(\ref{thetaweak}) and the strong-interaction Equations (\ref{strong}),(\ref{thetastrong}) separated by the condition $u=u_c$, constitute our main result. As shown in Fig.~\ref{GSSOPT}, they agree very well with the numerical optimization results. The weak-interaction power-laws $\sigma_o\propto j^{1/3}$, $s_o\propto j^{2/3}$ are continuously replaced by the $\sqrt{j}$ scaling of both quantities as the interaction strength is increased. It is also evident from Eq.~(\ref{strong}), that the optimal squeezing changes from initial number squeezing to initial phase-squeezing at $u\theta\approx 1$.  

%fig4
%%%%%%%%%%%%%%%%%%%%%%%
\begin{figure}[t]
\centering
\includegraphics[angle=0,width=0.5\textwidth]{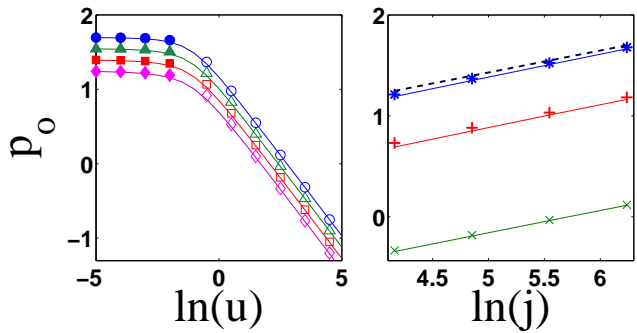}
\caption{(Color online) Optimal precision for a spin coherent state as a function of $u$ at fixed $j$ (left) and as a function of $j$ at fixed $u$ (right). Parameter values and notation are the same as in Fig.~\ref{GSSOPT} with symbols corresponding to Eq.~(\ref{sSCS}).}
\label{SCSopt}
\end{figure}
%%%%%%%%%%%%%%%%%%%%%%

%fig5
%%%%%%%%%%%%%%%%%%%%%%%
\begin{figure}[t]
\centering
\includegraphics[angle=0,width=0.5\textwidth]{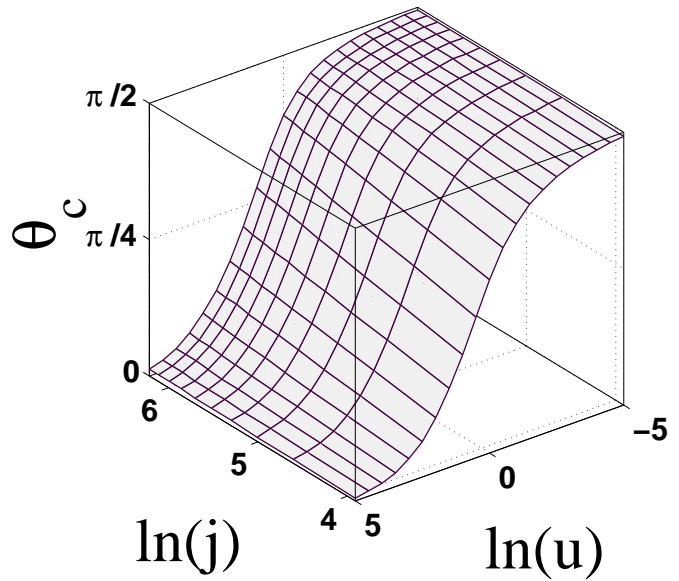}
\caption{(Color online) Same as Fig.~\ref{GSST} for spin coherent states.}
\label{SCST}
\end{figure}
%%%%%%%%%%%%%%%%%%%%%%

\section{Comparison with coherent input}
It is instructive to compare the optimized SNR with that of an initial spin coherent state $\exp[i\pi \Jy/2]|j,-j\rangle$, approaching the $\sigma=\sqrt{j/2}$ Gaussian, in the presence of interactions. Numerical optimization results are shown in Fig.~\ref{SCSopt} and Fig.~\ref{SCST}. The phase diffusion of the coherent state is given as,
\begin{eqnarray}
\langle\Jx(\theta)\rangle&=&j(\cos\tau)^{2j-1}~,\\
\left[\Delta J_x(\theta)\right]^2&=&\frac{j^2}{2}\left[ 1+(\cos 2\tau)^{2j-2}-2(\cos\tau)^{4j-2}\right]\nonumber\\
~&~&+\frac{j}{4}\left[1-(\cos 2\tau)^{2j-2}\right]~,\\
\left[\Delta J_y(\theta)\right]^2&=&\frac{j^2}{2}\left[ 1-(\cos 2\tau)^{2j-2}\right]\nonumber\\
~&~&+\frac{j}{4}\left[1+(\cos 2\tau)^{2j-2}\right]~,
\end{eqnarray}
where $\tau=(u/j)\theta$.
For $u=0$ the optimal phase is $\theta_c=\pi/2$, minimizing projection noise and resulting in the SNR $s_{c}=\pi\sqrt{j/2}$. In the presence of interactions $\theta_c$ (Fig~\ref{SCST}) decreases to reduce the phase-diffusion time. The optimal SNR is given by,
\begin{equation}
s_c=\theta_c\sqrt{2j}\left(1+4u^2\theta_c^2\right)^{-1/2},
\label{sSCS}
\end{equation}
with the optimal relative-phase given by,
\begin{equation}
2\theta_c^3\tan\theta_c\left(1+\tan^2\theta_c\right)=j/u^2~.
\end{equation}
The best SNR for a coherent preparation thus approaches $s_c\approx \sqrt{j/(2u^2)}$ for $u\theta_c\gg 1$.

%fig6
%%%%%%%%%%%%%%%%%%%%%%%
\begin{figure}[t]
\centering
\includegraphics[angle=0,width=0.5\textwidth]{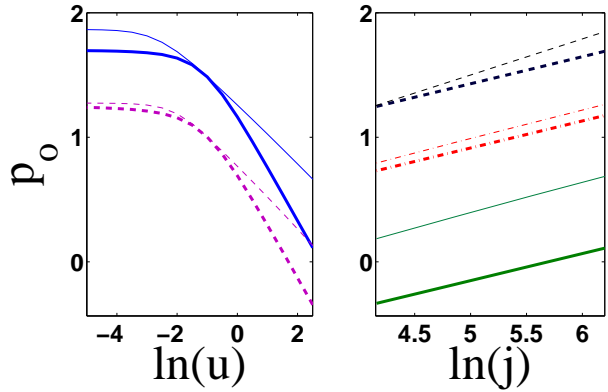}
\caption{(Color online) Comparison of best precision for Gaussian squeezed states (normal lines) vs. spin coherent states (bold lines). Left panel shows the dependence on $u$ at fixed $j=64$ (dashed)  and $512$ (solid).  Right panel depicts the dependence on $j$ for fixed $u=0$   and as a function of $j$ at fixed $u=0$ (dashed), $\ln(u)=0$ (dash-dotted), and $\ln(u)=2.5$ (solid). }
\label{GSSvsSCS}
\end{figure}
%%%%%%%%%%%%%%%%%%%%%%

In Fig.~\ref{GSSvsSCS} we compare the best precision obtained for spin coherent states (bold lines) with that of the optimized Gaussian squeezed states. For weak interactions, the coherent states preform worse than the initially number-squeezed states due to the larger projection noise, with the anticipated $j^{1/2}$ vs. $j^{2/3}$ respective signal-to-noise ratios. However, as the interactions increase, the coherent states are less affected by phase diffusion due to their smaller hold-time number variance.  The number squeezing of the optimal Gaussian state decreases with $u$ in order to slow down phase-diffusion, until at $u=u_c$ it coincides with the coherent state. Beyond this point, the optimal Gaussians are initially phase-squeezed, rotated to number-squeezed states during the phase acquisition, thus slowing down phase-diffusion with respect to the coherent states. While retaining the same $s_o\propto j^{1/2}$ scaling at fixed $u$, the best Gaussian states offer a factor of $\sqrt{u}$ improvement in the SNR over coherent states,  as evident from comparison of Eq.~(\ref{strong}) and Eq.~(\ref{sSCS}).

\section{Conclusions}
Interaction-induced phase-diffusion is currently the most prominent obstacle in the way of sub-shotnoise atom interferometry \cite{Grond10}.  By optimizing the SNR of a Mach-Zehnder atom-interferometer in the presence of interactions, we confirmed and quantified the notion that whereas phase-squeezing during the phase-acquisition time is required to go below the standard quantum limit, the robustness of number-squeezed states to phase-diffusion makes them the preferred choice when interactions are sufficiently large. The transition from optimal number-squeezing to optimal phase-squeezing takes place at $u\sim u_c \propto j^{-1/3}$. The scaling of the best SNR with $j$ changes from the sub-shotnoise interaction-free Eqs.~(\ref{sigmaNI}),(\ref{sNI}) \cite{Huang08} through the weak-interaction Eq.~(\ref{weak}) to the strong-interaction behavior of Eq.~(\ref{strong}).

Before closing, it is instructive to compare in detail our results in the strong-interaction regime with those of Ref.~\cite{Grond10}.  This work optimizes the useful squeezing with respect to $\sigma$ for any given acquired phase $\theta$. This approach amounts to neglecting the $\Delta J_x$ contributions and retaining only the $\sigma^2/j^2$ term in Eq.~(\ref{QNI}) and the $u^2\theta^2/\sigma^2$ term in Eq.~(\ref{QI}). The optimal $\sigma$ for any fixed value of $\theta$ is $\sigma_o=\sqrt{ju\theta}=j\sqrt{\tau}$, giving $Q_o=2u\theta/j=2\tau$ and $s_o=\theta/\sqrt{Q_o}=\sqrt{\theta j / (2u)}$. These expressions are the same as those appearing in \cite{Grond10}  and superficially seem to coincide with our Eq.~(\ref{strong}). However, checking for self-consistency by substituting $\sigma_o$ into (\ref{QNI}) and (\ref{QI}), we obtain that $(\Delta J_x)^2$ could be neglected with respect to $(\Delta J_y)^2$ only when $j^{1/2}<2\sigma_o<j^{2/3}$, because the $\Delta J_x$ variance grows for both number- and phase-squeezed states, whereas the $\Delta J_y$ increases monotonically with hold-time number-squeezing. This greatly restricts the validity of the useful-squeezing optimization to values of $\theta$ for which $\sigma_o$ lies in the appropriate squeezing window. By contrast, our calculation extends  to all values of $\theta$ and $\sigma_o$ by including also the $\Delta J_x$ contribution. 

In order to maximize the interferometer precision, it is imperative to devise schemes which will overcome or control phase-diffusion. One such approach may be to separate phase-acquisition from phase-diffusion, by reversing the roles of $\varepsilon$ and $K$, so as to measure frequency-shifts of Rabi oscillations. This will allow for the use of robust number-squeezed state without loosing readout accuracy, at the expense of having to follow an essentially nonlinear oscillation. Future work will also seek to similarly optimize an $SU(1,1)$ interferometer, based on the stimulated dissociation of molecular BECs \cite{Tikhonenkov09}. 

%%%%%%%%%%%%%%%%%%%%%%%%%%%%%%%%%%%%%%%%%%%%%%%%%%%%%%%%%%%%%%%%%%%%%%%%%%%%%%
%%%%%%%%%%%%%%%%%%%%%%%%%%%%%%%%%%%%%%%%%%%%%%%%%%%%%%%%%%%%%%%%%%%%%%%%%%%%%%
% conclusions

%%%%%%%%%%%%%%%%%%%%%%%%%%%%%%%%%%%%%%%%%%%%%%%%%%%%%%%%%%%%%%%%%%%%%%%%%%%%%%
%%%%%%%%%%%%%%%%%%%%%%%%%%%%%%%%%%%%%%%%%%%%%%%%%%%%%%%%%%%%%%%%%%%%%%%%%%%%%%

\begin{acknowledgements}
This work was supported  by the Israel Science Foundation (Grant 582/07), by grant no. 2008141 from the United States-Israel Binational Science Foundation (BSF), and by the National Science Foundation through a grant for the Institute for Theoretical Atomic, Molecular, and Optical Physics at Harvard University and Smithsonian Astrophysical Observatory.
\end{acknowledgements}

\end{document}